\documentclass[useAMS,usenatbib]{mn2e}

\usepackage{graphicx,times}

\def\aj{{AJ}}

\def\r1{$r_1$}
\title[1Gyr in the Life of NGC 6397]{1Gyr in the Life of the Globular Cluster NGC 6397}
\author[D.C. Heggie and M. Giersz]
{Douglas
  C. Heggie$^{1}$
  \thanks{E-mail:
 d.c.heggie@ed.ac.uk (DCH); mig@camk.edu.pl (MG)} and Mirek Giersz$^{2}$
\\
$^1$School of Mathematics and Maxwell Institute for Mathematical
Sciences, University of Edinburgh, King's Buildings, Edinburgh EH9
3JZ, UK\\
$^{2}$Nicolaus Copernicus Astronomical Centre, Polish Academy of Sciences, ul. Bartycka 18, 00-716 Warsaw, Poland\\
}
\begin{document}

\date{Accepted \ldots. Received \ldots; in original form \ldots}

\pagerange{\pageref{firstpage}--\pageref{lastpage}} \pubyear{2002}

\maketitle

\label{firstpage}

\begin{abstract}
M4 and NGC 6397 are two very similar galactic globular clusters, which
differ mainly in their surface brightness profile.  M4 has a classic
King-like profile, whereas NGC 6397 has a more concentrated profile,
which is often interpreted as that of a post-core collapse cluster.
\citet{HG2008}, however, found that M4 is also a post-core
collapse cluster, and \citet{GH2009} concluded that the main
reason for the difference between the two surface brightness profiles
is fluctuations.  This conclusion was reached on the basis of Monte
Carlo models, however, and in the present Letter we verify that
similar fluctuations occur in $N$-body models.  The models were initialised by
generating initial conditions from the Monte Carlo model of NGC6397 at the
simulated age of 12Gyr, and one was followed for 1Gyr.  The new models
help to clarify the nature of the fluctuations, which have the nature
of semi-regular oscillations with a time scale of order $10^8$yr.
They are influenced by the dynamical role which is played by
primordial binaries in the evolution of the core.
\end{abstract}


\section{Introduction}

A long-standing problem in the dynamics of galactic globular clusters
is the observed dichotomy in their surface brightness profiles.  While
most clusters exhibit a profile similar to a classic King profile
\citep{Ki1966}, about one quarter of the clusters exhibit a more cuspy
profile \citep{CD1989,Tr1995}.  The second class
of objects are usually described as post-core collapse clusters, as
the phenomenon of core collapse leads to an object with high central
density and small core radius \citep[etc.]{LBW1968,La1970,LBE1980}.  By a statistical study of stellar
luminosity functions, however, \citet{dempp2007} suggested
that some clusters with a King-like profile might, in fact, be
post-core collapse clusters.  Independently, \citet{HG2008} used a
Monte Carlo code for star cluster evolution to 
construct a dynamic, evolutionary model of the King-like
globular cluster M4, and were surprised to find that their model
was in the post collapse phase of its evolution at the present day.
They suggested that the reason it exhibited a finite core radius 
is that the core was sustained by heating from a population
of primordial binary stars.  (Many studies had shown that such a
population, certainly in the case of idealised models with stars of
equal mass, was sufficient to sustain the core in this way after 
core collapse; see, for example, \citet{MHM1990,HTH2006}.)

The case of the globular cluster NGC 6397 casts serious doubt on this
interpretation.  This cluster has a very similar mass to M4, and, if
anything, a larger population of primordial boundaries, and yet it
exhibits a non-King surface brightness profile.  If primordial
binaries sustain the finite core in M4, then NGC 6397 should exhibit a
finite core in the same way.  

This conundrum was considered by
\citet{GH2009}, who constructed a dynamic evolutionary model
for NGC 6397, just as they had previously done for M4.  Not
surprisingly, this model was also in its post-collapse evolution.
Despite having an appropriate primordial binary fraction, however, it
exhibited a non-King surface brightness profile that was a fair match
to that of NGC 6397.  They concluded that primordial binaries and core
collapse were insufficient to explain the surface brightness profiles
of these two clusters.  By considering and also eliminating other
alternatives, they concluded that the best explanation for the
difference between these clusters was one based on
fluctuations.  They found that, if they constructed the surface
brightness profile from a single model at different times, or from
different models started from the same initial conditions, but simply
with a different seed for the random number generator, then the
variety of surface brightness profiles exhibited differences
comparable to the difference between the surface brightness profiles
of M4 and NGC 6397.    In effect, their conclusion was the following:
that a single cluster could sometimes look like M4, and sometimes
look like NGC 6397.

Unfortunately it is not clear that a Monte Carlo code, such as the one
used for these clusters, simulates
fluctuations correctly.  \citet{GHH2008} did their
best to ensure that the Monte Carlo model behaves similarly to an
$N$-body model, at least in the range of $N$ where $N$-body models are
commonly carried out, but their investigation was restricted to global
quantities, such as the evolution of the total mass, and the half-mass
radius; they did not even consider the question of fluctuations.  The
nature of the fluctuations is also difficult to investigate with a
Monte Carlo code, which does not follow the orbital motions of the
particles.  These considerations motivate the $N$-body simulations
which we report in this letter.

The initial conditions we use are generated from our Monte Carlo model
of NGC 6397 evolved to the present day, which we took to be 12Gyr.  We placed the
model in a tidal field with a tidal radius equal to that of the Monte
Carlo model (which used a tidal cut-off), but switched off stellar
evolution.  One of the runs was continued for a simulation time of one
Gyr.  Details and results of the runs are given in the following two
sections, and the final section of the Letter summarises our conclusions.

\section{The simulations}


It is unfortunately impossible to specify the initial conditions
without access to the complete snapshot of the Monte Carlo model of
NGC 6397 at 12 Gyr, but table 1 gives some summary parameters.  The
Monte Carlo code stores the mass of every star, including binary components.
The only positional information on a star (or the barycentre of a
binary) which is held by the Monte Carlo code
is its radius, and the full position was generated with the assumption
of spherical symmetry.  For the velocity of each star (or barycentre) the
radial and transverse components are available, and the full velocity
vector was generated by assuming symmetry about the radial vector.
For a binary, the only internal information (apart from the binary
masses) { is for} the semi-major axis and eccentricity.  The full relative
position and velocity were generated assuming a random value of the
mean anomaly, and symmetric distributions of the orbital plane and the
line of apsides.

\begin{table}
\begin{center}
\caption{Data on the { initial} $N$-body model {(from a Monte Carlo model
  at 12Gyr)}}
\begin{tabular}{ll}
Number of single stars&105615\\
Number of binaries&3277\\
Total particle number&112169\\
Total mass &62000\\
Mass of binaries&2400\\
Mass of white dwarfs&27000\\
Mass of neutron stars&3100\\
Tidal radius&22\\
Half-mass radius&3.2\\
Core radius&0.05
\label{tab:ics}
\end{tabular}
 \end{center}
 Note: masses are given in $M_\odot$, radii in pc.  The ``core radius''
 is determined as in the $N$-body code (Aarseth 2003, pp.265-266).
 \end{table}


The simulation was run with NBODY6 on a PC equipped with a GPU.  The
CPU was a quad-core Intel Xeon E5410 at 2.33GHz, and the GPU a GeForce
9800 GTX.  As usual, the code uses $N$-body units \citep{HM1986}, but
the initial virial radius of the model (the $N$-body unit of length)
was 3.43pc, and its crossing time 1.08 Myr ($2\sqrt{2}$ $N$-body time
units).  The half-mass relaxation time is of order 700Myr. 

We actually carried out two runs.  One was an exploratory (but
scientifically informative) run with { dynamically} ``inert'' binaries, i.e. each
binary was replaced by a single particle with a mass equal to the
combined mass of the components.  This was run for an equivalent of
almost 260Myr.  The main run used { dynamically} ``active'' binaries, as described
above.    We refer to these two runs as ``I'' and ``A'' respectively.
Both simulations proceeded at a rate of
about 1Myr/hr, and the entire run with { dynamically} active binaries took about 1
month for 1Gyr.

\section{Results and Discussion}

\subsection{Structural evolution}

\begin{figure}
{\includegraphics[height=12cm,angle=0,width=9cm]{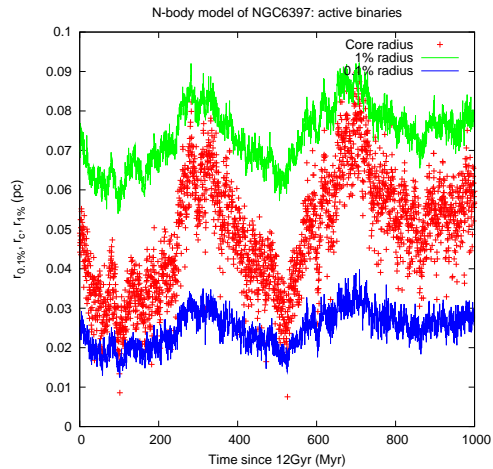}}
    \caption{Core and Lagrangian radii (mass fractions 0.1 and 1\%)
    for the model with { dynamically} active binaries { (Model A)}.  
    }
\label{fig:radii-active}
  \end{figure}

\begin{figure}
{\includegraphics[height=12cm,angle=0,width=9cm]{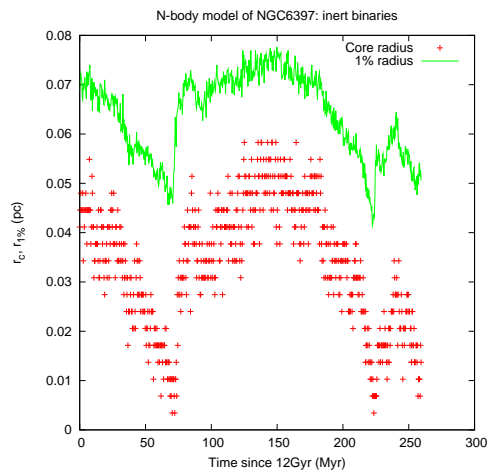}}
    \caption{Core and Lagrangian radius (mass fraction 1\%)
    for the model with { dynamically} inert binaries { (Model I)}.  
    Data on the 0.1\% Lagrangian radius are not available for this
    model, and the core radius was stored  only to 3 decimal places. }
\label{fig:radii-inert}
  \end{figure}

In the context set out in the Introduction, most of our interest is
focused on the inner parts of the models, and information on their spatial structure is
given in Figs.\ref{fig:radii-active} and \ref{fig:radii-inert} for the models
with { dynamically} active and inert binaries, respectively.  The core radius is as
defined in NBODY6 \citep[pp.265-6]{Aa2003}, and all radii are referred
to the density centre.

We note immediately substantial variations in \r1 (defined here to be
the 1\% Lagrangian radius).  Since the crossing time at this radius is
of order $10^4$yr, it is clear that these are not the kind of
fluctuations caused simply by the motion of stars in and out of the
core, but take place on a much longer time scale.  By comparing the
density within the half-mass radius and \r1, and the above estimate of
the half-mass relaxation time, we may estimate that the relaxation
time at \r1 is under 1 Myr (the unit of time in these figures), and
the variations of largest amplitude (e.g. the rise between 500 and 700
Myr in fig.\ref{fig:radii-active}) take place on a much longer time scale.  

The next obvious observation is that the long-term fluctuations have a
comparable amplitude for the two  runs.  But the maximum and minimum
radii are lower in run I.  In particular, the deep minima in  run I
are not found in run A.  It is natural to attribute these facts to the
presence of primordial binaries: run I has to make its own binaries,
which requires higher density than the burning of an existing binary.
 The amplitude of the fluctuations in the inner Lagrangian radii { is}
 smaller than { that of} the core radius, but not inconsiderable,
resulting in variations in the mean density within the 1\% Lagrangian
radius by a factor of around 3:1.

Though we have not shown data on the outer
Lagrangian radii, we find that 
the variations we observe are largely confined to the inner few
percent of the mass.    In fact the 10\% Lagrangian radius fluctuates with
a relative amplitude of order 7\%, and the oscillations are
approximately $180^\circ$ out of phase with those of the core.  The
relative amplitude of oscillations drops to less than 1\% at the
half-mass radius.

The initial decrease in both runs is an interesting feature.  One
might suspect that it is due to some property of the Monte Carlo code,
which perhaps maintains thermal equilibrium in a structure which would
not be in thermal equilibrium in an $N$-body model.  { Another
  possibility is a response of the system to the cessation of mass
  loss by stellar evolution.} And yet, in the overall
range of data in these figures, the structure at the start is not
unusual.  Furthermore, it must be remembered that this model was
selected because, at 12 Gyr, it yielded a surface brightness profile
resembling that of NGC6397, and this happens only intermittently, even
in the Monte Carlo model.

\begin{figure}
{\includegraphics[height=12cm,angle=0,width=9cm]{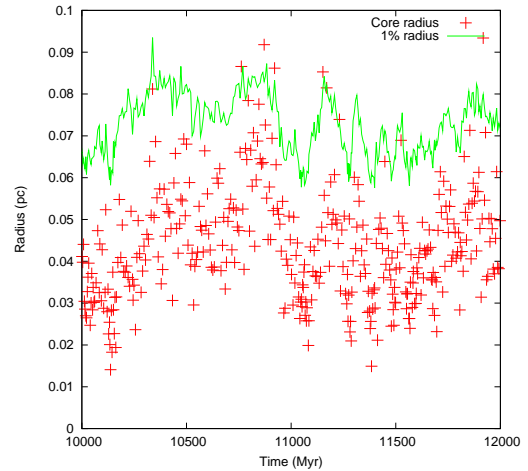}}
    \caption{Core and Lagrangian radius (mass fraction 1\%)
    for the Monte Carlo model (which has { dynamically} active binaries) between 10
    and 12 Gyr.  
    Data on the 0.1\% Lagrangian radius are not available for this
    model, and data were stored with lower frequency. The core radius
    $r_c$ is defined by 
$r_c^2 = {3\langle v^2\rangle}/({4\pi G\rho_0}),$
where the mass-weighted mean square three-dimensional velocity $\langle
v^2\rangle$ and the central density $\rho_0$ are calculated for the
innermost 20 stars.
}
\label{fig:mc-radii}
  \end{figure}

Fig.\ref{fig:mc-radii}, when considered in conjunction with
Figs.\ref{fig:radii-active} and \ref{fig:radii-inert}, allows a direct qualitative comparison between
the Monte Carlo and $N$-body models in respect of the fluctuations in
the inner radii, though the definition of the core radius adopted in
the Monte Carlo model is different (see caption).  This figure shows oscillations of
a similar amplitude and range of time scales to those exhibited by Run A
(Fig.\ref{fig:radii-active}). 

\begin{figure}
{\includegraphics[height=12cm,angle=0,width=9cm]{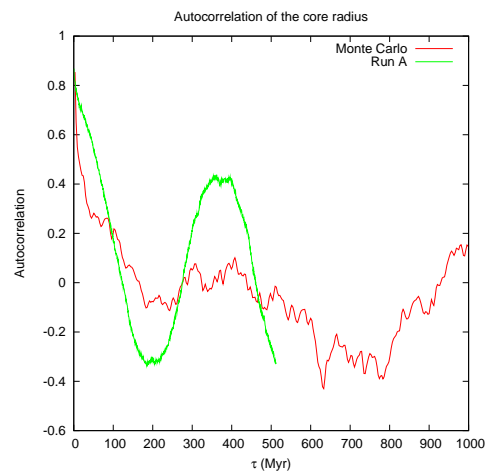}}
    \caption{Autocorrelation of the core radius in Run A and in the
      Monte Carlo model.  For the latter, which was output at
      irregular intervals, the data was interpolated.  The abscissa
      $\tau$ is defined in the text.
}
\label{fig:auto}
  \end{figure}

 In an attempt to make this statement more quantitative we show in
Fig.\ref{fig:auto} the autocorrelation of the core radius for Run A
and the Monte Carlo model, though it should be borne in mind that
these correspond to different time intervals.  In both cases we have
restricted the offset $\tau$ in the autocorrelation (defined to be
$\langle x(t)x(t+\tau)\rangle$ for a zero-mean normalised signal
$x(t)$) to be less than half the duration of the measurements.  The
result for the $N$-body model is striking, and confirms the visual
impression from Fig.\ref{fig:radii-active} of fairly regular
oscillations with a period of about 400Myr.  There is a faint
suggestion of similar structure in the autocorrelation function of the
Monte Carlo model.

Another structural quantity of interest is, of course, the total mass,
and the $N$-body run provides a check on this aspect of the Monte
Carlo model.  In the $N$-body model the rate of mass loss increases
for the first few tens of Myr, presumably because escapers take some
time to reach the boundary (at 2 tidal radii) where escapers are
removed.  For the period after 200 to 400Myr, however, the rate of mass loss
is virtually constant, and yields $d(\ln M)/dt = -1.35\times10^{-4}
\pm 2\times10^{-7}$, where the unit of time is 1Myr.  For the Monte
Carlo model discussed above, we do not have data extending beyond
12Gyr.  For another model differing only in the initial seed, however,
in
the same  period  the corresponding
result is $d(\ln M)/dt = -1.546\times10^{-4} \pm 5\times10^{-7}$.
This difference of 13\% should be corrected for the fact that
 there is no mass loss through stellar evolution in the $N$-body
model.   In the Monte Carlo model this contributes only about 3\% of the
total, but  the direct loss of this mass also induces further
loss of stars (by tidal overflow) because it make the potential well
more shallow.  It is difficult to quantify this induced mass loss, but
it appears that the discrepancy in the total rate of mass loss between
the two models is less than 10\%.

\subsection{Surface brightness profiles}

\begin{figure}
{\includegraphics[height=12cm,angle=0,width=9cm]{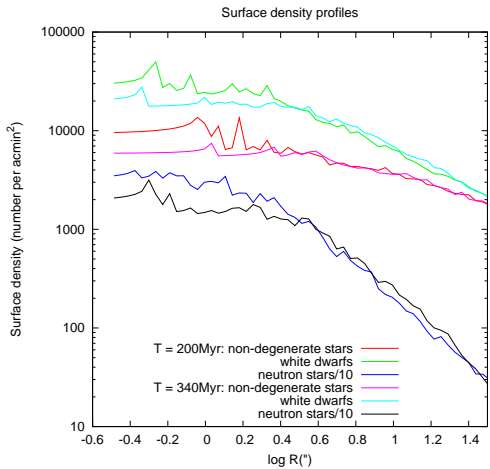}}
    \caption{Surface densities of three groups of stars in Run A at
      two times.  Only the central 30 arcsec (approximately) 
is shown.  For neutron
      stars the surface density is reduced by a factor 10, to avoid
      confusion with the surface density of white dwarfs.
 }
\label{fig:profiles}
  \end{figure}

Since so much of the mass at small radii is in the form of degenerate
remnants, the influence of the fluctuations shown in
Fig.\ref{fig:radii-active} on the surface brightness
profile is not obvious.  Insufficient data were collected from Run I,
but Fig.\ref{fig:profiles} provides a useful comparison of two snapshots
from Run A.  These are taken at times of 200 and 340 Myr, which
correspond, respectively, to low and high values of the radii plotted
in Fig.\ref{fig:radii-active} (though not the extreme values).  As
expected, the central surface densities are lower at the time when the
Lagrangian and core radii are larger.

Let us consider these profiles in more detail.  Neutron stars and
white dwarfs are the dominant components at the centre, each
contributing about half of the projected number density.  (Note that the
surface density of neutron stars in Fig.\ref{fig:profiles} is divided
by 10, to avoid overlap with the white dwarfs.)  Beyond the core,
however, the projected density of neutron stars decreases more
strongly, as is expected from their greater mass.  Indeed
they are actually the dominant central component in terms of {\sl spatial}
mass density.  The component which contributes least to the projected
central density, by a factor of order 10,
are the non-degenerate stars, though they begin to dominate at radii
beyond those shown in the figure.

All three components show a lower surface density and a larger core at
the later epoch (340 Myr) compared to the earlier epoch (200 Myr).  It
is interesting to attempt to quantify this in terms as close as
possible to observational procedures, by considering the radius at
which the projected density of the non-degenerate component drops below its central value by a factor
of two.  Despite the spatial fluctuations, it is clear that this radius is
about 15'' at the later epoch, and about 5'' at the earlier one.
Actually, it would also be reasonable to describe the latter profile
as a shallow cusp.  Indeed, in view of the very small number of stars
within 1 arcsec, this is a more robust description.  

At radii beyond about 4 arcsec the changes in the surface density are
much smaller but go in the opposite direction, at least for the
degenerate components.  At the epoch when  the core and Lagrangian
radii are larger, the number density at these radii is higher, the total mass of each
component being conserved (except for escape).

{ This discussion still does not necessarily correspond to an
observational description, which is usually expressed in terms of
surface brightness, i.e. weighted by luminosity.  In particular, the
most luminous stars are centrally concentrated relative to most
non-degenerate stars.  Because stellar evolution is absent from the
$N$-body simulation, however, it is not an entirely appropriate model
for considering how fluctuations affect the surface brightness.  The
discussion of our Monte Carlo model in \citet{GH2009} is more complete
in this respect.}

\subsection{Binaries}

\begin{figure}
{\includegraphics[height=12cm,angle=0,width=9cm]{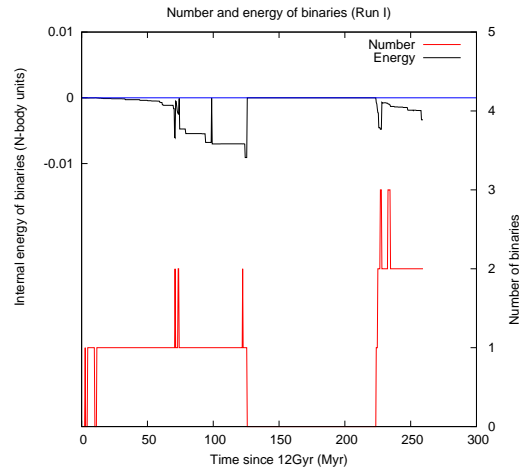}}
    \caption{Number and internal energy of binaries in run I.  The
 units of energy are $N$-body units \citep{HM1986}.
 }
\label{fig:binaries-I}
  \end{figure}

Run A contains over 3000 binaries, and it is best to begin with run
I, where the effects of individual binaries are clearer
(Fig.\ref{fig:binaries-I}).  By comparing with
Fig.\ref{fig:radii-inert}, we can surmise that one or two binaries
were responsible for terminating the core collapse at about 70Myr and
initiating the subsequent expansion.  For a period beginning at
125Myr, however, no binaries were present, suggesting that the
continuing modest expansion was powered gravothermally
\citep{SB1983}.  A second phase of core collapse and binary formation
appears to have started at about 225Myr, but is incomplete by the end
of the run.

  \begin{figure}
{\includegraphics[height=12cm,angle=0,width=9cm]{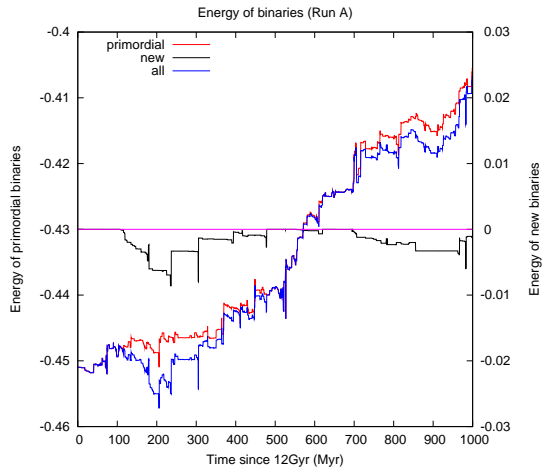}}
    \caption{ Internal energy of primordial and new binaries in run
 A.  Though the origins differ (because of the enormous energy of the
 primordial binary population) the scales on the vertical axes are
 equal.  The meaning of ``new'' and ``primordial'' is discussed in the text.
 }
\label{fig:binaries-A}
  \end{figure}

Fig.\ref{fig:binaries-A} shows similar data for run A.  The numbers
are not included, but the binaries are separated into two groups.
Those labelled ``new'' {\sl exclude} binaries which have evolved from
primordial binaries by exchange.  If, however, two binaries collide,
resulting in a hierarchical triple system, the outer motion of the
hierarchy is regarded as a ``new'' binary.
In any event we see
that both types of binary have an active role to play.  The
correlation with the evolution of the radii
(Fig.\ref{fig:radii-active}) is  less clear than for run I,
though the initial contraction seems to be associated with a period of
relatively sluggish binary activity.

While neither run shows the kind of gravothermal oscillations which
are so evident in simulations of systems with equal masses, it should
not be surprising to find evidence of  gravothermal effects in a
multi-mass system of the kind which we are studying (see the
discussion in \citet{GH2009}).

The rate of change of the binary fraction in run A is $+2.05\times10^{-7}\pm
2.5\times10^{-8}$, where the unit of time is 1Myr.  For the Monte
Carlo model the corresponding value  is $+1.56\times10^{-7}\pm
1.5\times10^{-8}$, and these results are pretty consistent within the
errors.

\section{Conclusions}

We have carried out an $N$-body simulation of the globular cluster
NGC6397, in order to study the evolution of its central structure over
a period of 1Gyr starting at the present day.  The
simulation was initialised using the results of a Monte Carlo model
\citep{GH2009} which approximately fits the present-day profiles of
surface brightness and velocity dispersion, and the mass function at
two radii.  The main limitation of the $N$-body model is that there is
no stellar evolution.  This apart, it is in many respects the most
realistic $N$-body simulation of a specific globular cluster of which
we are aware.

The model provides a new check on the reliability of the Monte Carlo
code, and suggests that the evolution of the binary fraction is
satisfactory, while the rate of escape of mass appears to be in agreement to
better than 10\%.

The results show that the population of primordial binaries in the
cluster (about 3\%) suppresses the deepest collapses of the core,
which nevertheless still exhibits substantial fluctuations on a time scale
of many core relaxation times.  Their amplitude is sufficient to
change the mean density of the innermost 1\% of the mass by a factor
of order 3:1.  These changes are reflected in variations in the core
radius, whether measured in terms of the spatial density, as in an
$N$-body model, or by the radius at which the surface density
decreases to half its central value.  These changes manifest
themselves in all components that we have studied: neutron stars (the
dominant component in the central density), white dwarfs, and
non-degenerate stars, which are the
least dominant.

In view of our results, it is interesting to think of globular
clusters as {\sl variable stellar systems}, in much the same way that
many stars are variable.  The time scales and mechanisms are vastly
different, and the variations in globular clusters are confined to the
vicinity of the core.  But it is another demonstration of the deep
physical resemblance between these two types of thermal,
self-gravitating objects.

\section*{Acknowledgements}

We are indebted to S. Aarseth and K. Nitadori for making publicly
available their version of NBODY6 adapted for use with a
GPU. S. Aarseth also helped specifically with aspects of input and
output.  Our hardware was purchased using a Small Project Grant
awarded to DCH and Dr M. Ruffert (School of Mathematics) by the
University of Edinburgh Development Trust, and we are most grateful
for it.  We thank S. Law and D. Marsh for installing and maintaining
it.  { We are grateful to the referee for his expeditious report.} 
This work was partly supported by Polish Ministry of Science and
Higher Education through the grant 92/N--ASTROSIM/2008/0.

\bsp

\label{lastpage}

\end{document}